%% file: conference_101719.tex
\documentclass[conference]{IEEEtran}
\IEEEoverridecommandlockouts
\usepackage{cite}
\usepackage{amsmath,amssymb,amsfonts}
\usepackage{algorithmic}
\usepackage{graphicx}
\usepackage{textcomp}
\usepackage{xcolor}
\usepackage{cite}
\usepackage{amsmath,amssymb,amsfonts}
\usepackage{algorithmic}
\usepackage{graphicx}
\usepackage{textcomp}
\usepackage{xcolor}
\usepackage{blindtext}
\usepackage{float}
\usepackage{tikz}
\usepackage{amsmath}
\usetikzlibrary{patterns}

\usepackage{filecontents}
\usepackage{graphicx}
\usepackage{caption}
\usepackage{subcaption}

\usepackage{tikz}
\usepackage{amsmath}
\usepackage{filecontents}
\usepackage{framed}
\usepackage{color}
\usepackage{xcolor}
\definecolor{LimeGreen}{rgb}{0.2, 0.8, 0.2}
\usepackage[linesnumbered,ruled,vlined]{algorithm2e}

\usepackage{algorithmic}
\usepackage{array}
\usepackage{multirow}
\usepackage{booktabs,subcaption,amsfonts,dcolumn}
\usepackage{paralist}
\usepackage{balance}
\usepackage{todonotes}
\usepackage{hyperref}
\usepackage{wrapfig}
\usepackage{wrapfig}
\usepackage [ n,
advantage , operators , sets , adversary , landau , probability , notions , logic ,
ff, mm, primitives , events , complexity , asymptotics , keys]{ cryptocode}
\usepackage{bbding}
\usepackage{dashrule}
\usepackage{tikz}
\usepackage{amsmath}
\usepackage{pifont}
\usepackage{multicol}
\usepackage{amsthm}

\usepackage{filecontents}
\usepackage{tablefootnote}

\usepackage{bm}

\input{macros}
\usepackage{amsmath}
\usepackage{graphicx}

\usepackage{fancyhdr}
\usepackage{threeparttable}
\usepackage{lipsum}
\usepackage{enumitem}
\usepackage{pgfplots}
\pgfplotsset{compat=1.11,
    /pgfplots/ybar legend/.style={
    /pgfplots/legend image code/.code={%
       \draw[##1,/tikz/.cd,yshift=-0.25em]
        (0cm,0cm) rectangle (10pt,0.8em);},
   },
}
\usepgfplotslibrary{groupplots}
\usepackage[skip=-1pt]{caption}
\usepackage{flushend}
\usepackage{tikz}
\usepackage{amsmath}
\usepackage{xcolor}
\usepackage{titlesec}
\def\BibTeX{{\rm B\kern-.05em{\sc i\kern-.025em b}\kern-.08em
    T\kern-.1667em\lower.7ex\hbox{E}\kern-.125emX}}
\begin{document}

\title{Privacy-preserving Chunk Scheduling in a BitTorrent Implementation of Federated Learning%
\thanks{This paper has been accepted to the 46th IEEE International Conference on Distributed Computing Systems (ICDCS 2026). Please cite the IEEE proceedings version once it becomes available.}
}

\author{\IEEEauthorblockN{ Naicheng Li}
\IEEEauthorblockA{\textit{IMDEA Networks Institute} \\
Madrid, Spain \\
naicheng.li@networks.imdea.org}
\and
\IEEEauthorblockN{Javad Dogani}
\IEEEauthorblockA{\textit{IMDEA Networks Institute} \\
Madrid, Spain \\
javad.dogani@networks.imdea.org}
\and
\IEEEauthorblockN{Rui Wang}
\IEEEauthorblockA{\textit{Department of Software Technology} \\
\textit{Delft University of Technology}\\
Delft, The Netherlands \\
r.wang-8@tudelft.nl}
\and
\IEEEauthorblockN{Kaitai Liang}
\IEEEauthorblockA{\textit{Department of Computing \& Department of Intelligent Systems} \\
\textit{University of Turku \& Delft University of Technology}\\
Turku, Finland \& Delft, The Netherlands\\
kaitai.liang@utu.fi}
\and
\IEEEauthorblockN{Nikolaos Laoutaris}
\IEEEauthorblockA{\textit{IMDEA Networks Institute} \\
Madrid, Spain \\
nikolaos.laoutaris@networks.imdea.org}
}

\maketitle

\begin{abstract}
Traditional federated learning (FL) relies on a central aggregator server, which can create performance bottlenecks and privacy risks.
Decentralized \emph{mix-and-forward} designs remove the server, but repeated local mixing can attenuate global information under heterogeneity and exposes peer-to-peer neighborhoods as a privacy attack surface.
To preserve FedAvg-style aggregation semantics (over updates reconstructable by the round deadline) while scaling dissemination, we present \textbf{\emph{FLTorrent}}, a BitTorrent-based dissemination layer for serverless FL with a short warm-up.
Warm-up hardens \emph{within-round source unlinkability}---a dissemination-layer goal orthogonal to content protections (e.g., DP or secure aggregation)---via (i) pre-round obfuscation, (ii) randomized lags, and (iii) coordination-only non-owner-first scheduling (tracker off the data path), before switching to vanilla BitTorrent swarming.
We upper-bound the per-transfer attribution posterior by the fraction of owner chunks in a sender's eligible cover set, and derive a tighter high-probability bound that improves with early non-owner mass.
A simple heuristic, \textsc{GreedyFastestFirst}, attains $\approx 92\%$ of a bandwidth-optimal max-flow upper bound, while warm-up remains a stable $\approx 12\%$ share of a round across $100$--$500$ peers.
Under an observation-only local adversary, FLTorrent drives attribution success close to neighborhood-level random guessing for typical nodes, improves with network size, and remains robust under collusion.
In LLM-scale dissemination stress tests over 7--10 Gbps access links, FLTorrent adds only $\sim$6--10\% round-time overhead relative to BitTorrent-only.
Overall, FLTorrent shows that within-round unlinkability and BitTorrent-level efficiency can co-exist with predictable, low overheads at scale.
\end{abstract}


\input{Sections/introduction}

\input{Sections/3problem_formulation}

\input{Sections/4FLTorrent_Design}

\input{Sections/analysis}

\input{Sections/5ATTACKS_AGAINST_FLTORRENT}
\input{Sections/6experiment}

\input{Sections/7Related_work}
\input{Sections/8conclusion}

\input{Sections/Acknowledgment}

\bibliographystyle{ieeetr}
\bibliography{ref_short}


\end{document}

%% file: macros.tex
\newcommand{\node}{v}

%% file: Sections/Introduction.tex
\section{Introduction}

Federated Learning (FL) trains shared models without centralizing training data by exchanging model updates~\cite{mcmahan2017communication}.
In the canonical server-based workflow, a central aggregator collects client updates, computes a weighted average (FedAvg), and redistributes the new model~\cite{mcmahan2017communication}.
While widely adopted for privacy and compliance (e.g., GDPR), this architecture has its downsides: the server can become a communication bottleneck and requires full trust.
These issues have motivated substantial distributed-systems work on communication-efficient aggregation, client selection, personalization, and robustness~\cite{luo2023incentive,wu2023fedgnn_privacy,zhang2025fedlth,xia2024gain}.

Despite keeping data local, FL is not immune to privacy inference.
Prior work shows that adversaries can extract sensitive information from \emph{update contents} via membership inference and gradient inversion/reconstruction~\cite{shokri2017membership,zhu2019deep,geiping2020inverting,diana2025cuttingprivacyhyperplanebaseddata}.
Accordingly, many deployments adopt content-level defenses such as secure aggregation (and often DP), which limit visibility of individual updates to the aggregator~\cite{bonawitz2017practical,pasquini2022eluding,ngo2024secureaggregationprivatemembership}.
However, such defenses do not address \emph{metadata leakage} in the communication substrate---e.g., which participant likely produced which update and when---which becomes salient in peer-to-peer dissemination.
This motivates reducing reliance on a single trusted aggregator \emph{and} hardening the dissemination layer against within-round attribution.

A prominent direction is \emph{decentralized learning} and decentralized FL (DL/DFL), where clients exchange information over an overlay graph and update their models via local mixing (e.g., push-sum/gossip, mixing matrices, or ``mix-and-forward'' propagation)~\cite{assran2019stochastic,koloskova2020unified,yuan2024decentralized}.
This improves fault tolerance and reduces reliance on one aggregation point, but it complicates the ``what is being computed'' question and introduces proximity-driven privacy risks.
First, local mixing is inherently a multi-hop approximation: within a finite time budget, each node only sees \emph{partially mixed} information rather than the full per-round set of updates.
Under heterogeneity (e.g., non-IID data, bandwidth skew, and churn), such repeated mixing can distort the effective aggregation and degrade convergence/robustness, which we refer to as \emph{attenuation}.
Second, peer-to-peer (P2P) proximity changes the privacy attack surface: neighbors can observe fine-grained protocol signals (e.g., timing and early-transfer patterns) and exploit proximity-driven privacy risks in decentralized ML settings~\cite{pasquini2023security}.
Thus, existing decentralized designs often trade off among (i) FedAvg-style aggregation semantics, (ii) scalable dissemination, and (iii) robustness to proximity-driven privacy threats.

\vspace{2pt}
\noindent \textbf{Main idea.}
To avoid semantic attenuation while scaling dissemination, we build the communication substrate on BitTorrent~\cite{cohen2003incentives}.
Rather than repeatedly mixing models over hops, we treat each client's update as a file to be disseminated: the update is split into fixed-size chunks, and peers exchange chunks in parallel within a swarm.
Swarming amortizes upload cost (others replicate what you seed), leverages heterogeneous bandwidth, and sustains high utilization via tit-for-tat incentives~\cite{cohen2003incentives,cuevas}.
Most importantly, \emph{if each round's updates are fully disseminated (i.e., reconstructable) by the round deadline}, then every client locally computes the \emph{same} FedAvg aggregate as in server-based FL, without a central aggregator~\cite{mcmahan2017communication}.
When some updates are not reconstructable by the deadline (e.g., due to partial participation or timeouts), FLTorrent computes FedAvg over the reconstructable active set, matching the standard FL treatment of missing updates.
In other words, BitTorrent aligns dissemination with aggregation semantics: once peers obtain the same set of per-round updates by the deadline, they naturally agree on the same aggregate.

However, vanilla BitTorrent optimizes throughput and availability, not privacy.
In P2P FL, this mismatch creates a protocol-specific privacy objective: \emph{within-round source unlinkability} (attribution resistance) between a client and its update at the P2P layer.
Concretely, at round start, each client initially holds only its own chunks; consequently, early transmissions are biased toward ``owner'' chunks, and a neighbor can attribute an update by correlating which chunks it first observes from which sender.
Such attribution is not merely cosmetic: once an update can be linked to a specific client or silo (e.g., a hospital), the consequences become materially worse and can be actionable (e.g., revealing spatio-temporal operational signals).
For example, \emph{source inference attacks} aim to identify \emph{which client} contributed a particular training record; Hu \emph{et al.}~\cite{hu2021sourceinference} show that identifying the \emph{source hospital} can already be actionable (e.g., revealing whether a hospital is in a high-risk region), and Chen \emph{et al.}~\cite{ijcai2025p536} discuss that spatio-temporal source inference can reveal sensitive dynamics (e.g., during a rare disease outbreak).
Even when secure aggregation hides individual updates from the server, it does not preclude inference about participant-level attributes (e.g., client quality) from aggregated outputs~\cite{pejo2022quality}.
In this paper, we focus on local, observation-only adversaries within the swarm and on permissioned settings where admission control mitigates Sybil attacks; global traffic analysis is out of scope.

\vspace{2pt}
\noindent \textbf{Contributions.}
We introduce \textbf{FLTorrent}, a BitTorrent-based dissemination layer that preserves FedAvg-style aggregation over deadline-reconstructable updates while hardening P2P dissemination against \emph{within-round source attribution}.
FLTorrent first runs a short warm-up that injects uncertainty through (i) \emph{pre-round obfuscation}, (ii) randomized start lags, and (iii) tracker-coordinated non-owner-first scheduling, with the tracker remaining off the data path; it then returns to vanilla BitTorrent swarming.
Warm-up ends once each active peer has accumulated a global cover set of at least $k_\beta=\lceil\beta|\mathcal{C}^r|\rceil$ chunks, yielding an attribution bound that decreases with the required non-owner mass $h_u=\max\{0,k_\beta-K_u^r\}$.
We formalize this P2P attribution objective, derive unlinkability bounds, and characterize a max-flow-style upper bound together with practical warm-up scheduling heuristics.

\vspace{2pt}
\noindent \textbf{Results.} Our greedy warm-up scheduler, \textsc{GreedyFastestFirst}, attains $\approx 92\%$ of the max-flow-style upper bound.
Across 100--500 peers, warm-up remains a stable $\approx 11.5\%$--$12.4\%$ fraction of a round with overall utilization $75\%$--$80\%$.
With all defenses enabled, the maximum attribution success rate approaches random guessing (approximately $1/m$ under default connectivity).
In LLM-scale stress tests (Gemma-7B, DeepSeek-R1-14B, Qwen2.5-32B, Llama-3.3-70B) over 7--10\,Gbps links, FLTorrent adds only $\sim 6\%$--$10\%$ round-time overhead relative to BitTorrent-only.

%% file: Sections/3problem_formulation.tex
\section{Problem Formulation}\label{sec:problem}

\subsection{Background: BitTorrent primitives and FL rounds.}
BitTorrent~\cite{cohen2003incentives} disseminates an object by splitting it into fixed-size chunks (BitTorrent \emph{pieces}) and exchanging them in parallel among peers.
A \emph{torrent descriptor} (metadata) enumerates chunk identifiers (piece indices) and cryptographic hashes, enabling receivers to verify chunk integrity and discard corrupted payloads~\cite{bep0003}.
Peers advertise availability via \emph{bitfields} and use swarming incentives (e.g., tit-for-tat) to sustain high utilization~\cite{cohen2003incentives}.
In vanilla BitTorrent, the tracker mainly supports peer discovery and does not schedule chunks.
In \emph{FLTorrent}, the tracker additionally collects \emph{per-peer chunk availability} (bitfields) during a short warm-up period and issues early scheduling directives; it is not on the data path and does not need access to chunk contents.
A BitTorrent \emph{peer} corresponds to an FL \emph{client}; we use ``client'' throughout and use ``peer'' only when referring to BitTorrent mechanisms.

\subsection{System Model (One Round)}\label{sec:system-model}
We consider a BitTorrent-style tracker $\mathcal{T}$ and a set of clients $\mathcal{V}$ ($|\mathcal{V}|=n$).
For a single FL round $r\in\mathbb{N}$, the tracker forms an overlay graph $G^r=(\mathcal{V},\mathcal{E}^r)$ and coordinates warm-up scheduling using the bitfields it collects.
For $v\in\mathcal{V}$, its neighborhood is $\mathcal{N}^r(v)=\{w:(v,w)\in\mathcal{E}^r\}$ and the average degree is
$m=\frac{1}{n}\sum_{v\in\mathcal{V}}|\mathcal{N}^r(v)|$.
Time is slotted with index $s\in\{0,1,\dots\}$ and duration $\Delta=1$. Each round has a deadline at slot $s_{\max}$ (used throughout for warm-up termination and aggregation).
Each client $v$ has uplink/downlink capacities $U_v,D_v$ (bytes/slot).
Under a dissemination/scheduling \emph{policy} $\pi$, instantaneous rates $u_v[s;\pi]$ and $d_v[s;\pi]$ satisfy
$0\le u_v[s;\pi]\le U_v$, $0\le d_v[s;\pi]\le D_v$, and flow conservation
$\sum_{v\in\mathcal{V}} u_v[s;\pi]=\sum_{v\in\mathcal{V}} d_v[s;\pi]$ for all $s$.
When convenient, we convert byte budgets into per-slot chunk budgets as
$u_v=\lfloor U_v\Delta/C\rfloor$ and $d_v=\lfloor D_v\Delta/C\rfloor$.

Client $v$ produces update $g_v^r$ of size $S_v^r$ bytes, sliced into $K_v^r=\lceil S_v^r/C\rceil$ chunks of $C$ bytes each.
For exposition, we label these chunks as
$\mathcal{C}_v^r=\{(v,r,i):i=1,\dots,K_v^r\}$ and
$\mathcal{C}^r=\bigcup_{v\in\mathcal{V}}\mathcal{C}_v^r$.
(These tuples are \emph{analysis labels}; in the actual protocol, peers exchange a torrent identifier and a piece index, which do not explicitly encode $v$.)
For any client $v \in \mathcal{V}$, let $\mathcal{C}_v^r[s]\subseteq \mathcal{C}^r$ denote the chunks held by $v$ at slot $s$.
The (full) completion time under policy $\pi$ (assuming all clients remain active) is
$T_{\mathrm{full}}(\pi)=\min\{s:\forall v\in\mathcal{V}, \mathcal{C}_v^r[s]=\mathcal{C}^r\}$.


\noindent\textbf{Aggregation semantics and descriptor authentication.}
At round start, each active client registers exactly one torrent descriptor and scalar weight (e.g., local sample count).
In permissioned deployments, this registration is authenticated under the client's membership credential; the tracker publishes a signed per-round manifest containing descriptor hashes, scalar weights, and round pseudonyms.
Clients verify this manifest before downloading, and duplicate submissions under the same credential are rejected during registration.
In our analysis, we assume homogeneous update sizes across clients (i.e., $S_v^r = S$ for all $v$ and thus $K_v^r=K$); consequently, torrent descriptors---which contain only chunk hashes and piece counts---do not reveal owner identity.\footnote{When update sizes vary significantly, an attacker could narrow down candidate owners by matching observed chunk counts to known update sizes. Mitigations include padding updates to a common size or grouping clients with similar update sizes; we leave such extensions to future work.}
Peers learn only which chunks form a given update and its associated weight, not which client produced it.
By the round deadline $s_{\max}$, each client $v$ reconstructs the subset of updates whose chunks it has obtained and locally computes a FedAvg aggregate over its \emph{reconstructable set}
\[
g_{v}^{\text{agg},r}=\sum_{u\in\mathcal{A}_v^r}\tfrac{w_u}{\sum_{j\in\mathcal{A}_v^r}w_j}\,g_u^r, 
\mathcal{A}_v^r:=\{u\in\mathcal{V}:\mathcal{C}_u^r\subseteq \mathcal{C}_v^r[s_{\max}]\}
\]
with $|\mathcal{A}_v^r|\ge 1$ required for aggregation.
(Indexing by client identity $u$ is for analysis; in the protocol, peers identify updates by their published descriptors/pseudonyms, not by real identities.)
Under within-round dropouts, $\mathcal{A}_v^r$ excludes updates that $v$ cannot reconstruct by $s_{\max}$; a client may disconnect after seeding chunks, yet its update can still belong to $\mathcal{A}_v^r$ if chunks have been sufficiently replicated.

\noindent\textbf{Observations.}
Let $\mathcal{O}_v[s]$ denote client $v$'s observations up to slot $s$: received piece indices (chunk identifiers), sender round-pseudonyms, timestamps/order, and (during warm-up) tracker coordination messages it receives.
Round pseudonyms are \emph{stable within a round} (accounting/tit-for-tat) and \emph{rotate across rounds} to mitigate cross-round linkability.

\subsection{Threat Model}\label{sec:threat-model}

\noindent\textbf{Goal.}
We target \emph{peer-side within-round source unlinkability} (\S\ref{sec:privacy-metrics}) against inference from proximity-visible metadata.
Because clients publish torrent descriptors, the tracker inherently learns the chunk-to-owner mapping; hiding it from the tracker is \emph{not} our objective.

\noindent\textbf{Channels and scope.}
We assume secure and authenticated channels (e.g., TLS), which protect contents but not traffic metadata (endpoints/timing/volume); thus Internet-scale traffic analysis remains out of scope.
We assume round pseudonyms are tracker-issued and authenticated over these channels; spoofing another client's pseudonym reduces to credential compromise or Sybil admission and is out of scope.

\noindent\textbf{Adversary A (HBC peers).}
A static attacker corrupts a subset of clients before a round begins.
Corrupted clients follow the protocol but infer attribution from
(i) chunk indices received from each neighbor (identified by sender pseudonym),
(ii) transfer timing/ordering, and (iii) warm-up directives received from the tracker.
They may collude by pooling observations. Our attacks in \S\ref{attack} instantiate this adversary.

\noindent\textbf{Adversary B (Byzantine peers).}
We additionally consider Byzantine clients that may deviate arbitrarily (e.g., lie in bitfields, selectively forward/withhold chunks, or delay transmissions).
Payload tampering is detectable via chunk-hash verification from the torrent descriptor~\cite{bep0003};
thus undetectable content manipulation is out of scope.
Our unlinkability bounds (\S\ref{sec:probability}) apply to transfers \emph{sent by honest peers}; liveness under deviations is handled operationally (\S\ref{sec:fault-tolerance}).

\noindent\textbf{Tracker integrity assumption.}
The tracker coordinates overlay formation and warm-up directives but is not on the data path.
We assume a trusted tracker by default; an optional auditable (accountability) variant is described in \S\ref{sec:auditability}.
Tracker--client collusion is excluded since the tracker knows the chunk-to-owner mapping from descriptors and collusion would trivially reveal attribution.

\noindent\textbf{Out of scope.}
(i) Global network adversaries (traffic-pattern analysis; defense requires anonymous routing, orthogonal to our focus).
(ii) Sybil attacks (we assume permissioned membership).
(iii) Denial-of-service (e.g., forcing timeouts/fail-open) beyond our operational handling.
(iv) Update-content attacks (complementary to secure aggregation/DP).

\subsection{Privacy Objective and Metrics}\label{sec:privacy-metrics}
\noindent\textbf{Within-round source unlinkability.}
Under the above adversaries, BitTorrent-style dissemination introduces an attribution channel: early in a round, a sender is more likely to transmit its \emph{own} chunks, enabling neighbors to infer sender--source associations.
For any observed transfer of chunk $c$, let $\mathsf{snd}(c)$ be the sender (round-pseudonym) and $\mathsf{src}(c)$ the true source client of $c$.
We say the system attains \textbf{unlinkability level $P$ at slot $s$} if for all observers $v$ and observed transfers,
$\Pr[\mathsf{src}(c)=\mathsf{snd}(c)|\mathcal{O}_v[s]]\le 1/P$.
Let $T_P(\pi)=\min\{s:\text{unlinkability level $P$ holds}\}$ be the time to reach level~$P$.



\noindent\textbf{Warm-up knob $\beta$ (global cover-set threshold).}
FLTorrent uses a short warm-up phase and switches to vanilla BitTorrent once each active peer has accumulated a sufficiently large cover set.
Rather than parameterizing this threshold by the chunk count of a single update, we define it as a fraction of the swarm-wide chunk universe.
Let
\[
|\mathcal C^r|=\sum_{u\in V}K_u^r
\]
denote the total number of chunks published in round $r$.
Given a global cover ratio $\beta\in(0,1)$, FLTorrent sets
\[
k_\beta=\left\lceil \beta |\mathcal C^r| \right\rceil .
\]
Warm-up terminates once every active peer $v$ holds at least $k_\beta$ chunks in total:
\[
|\mathcal C_v^r[s]| \ge k_\beta,\quad \forall v\in V_{\mathrm{active}}^r .
\]
Since peer $v$ initially holds its own $K_v^r$ chunks, this requires it to acquire at least
\[
\max(0,k_\beta-K_v^r)
\]
non-owner chunks before the switch.
Under homogeneous update sizes, $K_v^r=K$ for all $v$, so
\[
k_\beta=\left\lceil \beta nK \right\rceil .
\]
For example, with $n=50$ clients, $K=10$ chunks per update, and $\beta=10\%$, each peer switches after holding $k_\beta=50$ chunks in total, i.e., after receiving at least $40$ non-owner chunks.
Our analysis (Section~\ref{sec:probability}) relates this global cover-set threshold to an attribution-posterior cap: for any honest sender, the posterior decreases as the required non-owner mass grows.

\noindent\textbf{Efficiency metrics.}
Over $H$ slots, aggregate throughput and normalized utilization are
\begin{gather*}
\textstyle
\bar{U}(\pi;H)=\frac{1}{H}{\sum}_{s,v} u_v[s;\pi],\\[2pt]
\textstyle
\mathrm{Util}(\pi;H)=\bar{U}(\pi;H)\big/{\sum}_v U_v\in[0,1].
\end{gather*}

\noindent\textbf{Goal.}
We seek policies $\pi$ that (i) reach a target unlinkability level quickly (small $T_P$), (ii) maintain high utilization, and (iii) minimize completion time $T_{\mathrm{full}}$.

%% file: Sections/4FLTorrent_Design.tex
\section{\emph{FLTorrent} Design}\label{sec:design}
We present \emph{FLTorrent} as a BitTorrent-based dissemination framework for serverless aggregation, with a warm-up phase that hardens the dissemination layer against within-round source attribution.
Section~\ref{sec:problem} defines the system model, threat model, and unlinkability objective; this section focuses on how it works.

\subsection{Overview and Round Workflow}\label{sec:protocol}
\emph{FLTorrent} runs in synchronous FL rounds. In each round $r$, clients disseminate their local updates via a BitTorrent-style swarm, with an explicit warm-up that hardens the dissemination layer against \emph{within-round source attribution}.
%
Each round consists of:

\noindent(1)~\textbf{Local training}---each client $v$ computes a local update $g_v^r$ from its private data;

\noindent(2)~\textbf{Chunking \& metadata publication}---$v$ splits $g_v^r$ into fixed-size chunks and publishes a torrent descriptor $\mathsf{desc}_v^r$ (with per-chunk hashes~\cite{bep0003}) and a round pseudonym $\mathsf{pid}_v^r$ (stable within-round, rotated across rounds);

\noindent(3)~\textbf{Warm-up}---the tracker coordinates early transfers so each active client accumulates at least $k_\beta=\lceil\beta|\mathcal{C}^r|\rceil$ chunks before normal BitTorrent announcements dominate;

\noindent(4)~\textbf{BitTorrent swarming}---clients switch to vanilla BitTorrent to fetch remaining chunks;

\noindent(5)~\textbf{Aggregation}---by the deadline, each client performs FedAvg over its reconstructable set $\mathcal{A}_v^r$ (Section~\ref{sec:system-model});

\noindent(6)~\textbf{Audit (optional)}---under the auditable tracker model, the tracker reveals per-round randomness and logs for post-hoc verification (Section~\ref{sec:auditability}).

We parameterize the warm-up threshold by the global cover ratio
$\beta$, i.e., $k_\beta=\lceil \beta|\mathcal C^r|\rceil$,
consistent with Section~\ref{sec:privacy-metrics}.

\noindent\textbf{Algorithm 1 (FLTorrent round workflow).}
The six steps above define a round: authenticated descriptor publication, overlay sampling, optional pre-round spray, scheduled warm-up until all active clients reach $k_\beta$, vanilla BitTorrent swarming, and deadline FedAvg over the reconstructable set $\mathcal{A}_v^r$.

\subsection{Warm-Up Phase for Source Unlinkability}\label{sec:warmup}
\noindent\textbf{Why warm-up is necessary.}
At round start, each client holds only its own local chunks.
If clients immediately request a vanilla BitTorrent announcement, early transfers are biased toward ``owner to neighbors,'' making attribution easy.
Warm-up mitigates this by ensuring each sender has a sufficiently large \emph{cover set} before normal BitTorrent dynamics dominate.


\noindent\textbf{Warm-up knob $\beta$ and unlinkability level.}
As defined in Section~\ref{sec:privacy-metrics}, warm-up terminates once each active client holds at least $k_\beta=\lceil\beta|\mathcal C^r|\rceil$ chunks.
The mechanisms below realize this global cover-set condition while reducing early owner-biased transfers; Section~\ref{sec:probability} formalizes how the resulting non-owner mass lowers the per-transfer attribution posterior.


\noindent\subsubsection{Pre-round Obfuscation Phase}
Pre-round obfuscation occurs before the round begins and injects a small amount of non-owner chunk mass into the system.
We parameterize its strength by a \emph{ratio} $R\in(0,1)$ (or $R$ in \%): for each source client $v$ with $K_v^r$ chunks, the tracker randomly selects
$\lfloor R\cdot K_v^r \rfloor$ \emph{chunk identifiers} from $\mathcal{C}_v^r$ and assigns each selected identifier to a randomly chosen \emph{non-neighbor} recipient.
To execute a spray transfer, the tracker sends control instructions (containing the chunk identifier) to both the owner $v$ and the designated recipient, directing them to establish an ephemeral direct tunnel, perform exactly one chunk transmission, and then tear the tunnel down.
These spray transfers use one-off spray pseudonyms that are not reused as the sender's round pseudonym during warm-up or BitTorrent swarming, and they are not announced to overlay neighbors.
Thus, aside from the tracker and the designated recipient, other peers do not observe the spray; the recipient learns only the spray pseudonym and piece identifier, not a stable sender identity.
Crucially, the tracker operates only on chunk identifiers and client identifiers---it never receives or forwards chunk payloads, and observes completion acknowledgments.
We choose a small $R$ (default $R=0.2$) to balance the injected non-owner mass against control-plane coordination overhead.

\noindent\subsubsection{Time Obfuscation}\label{sec:time-obf}
Pre-round obfuscation alone is insufficient if all clients start transmitting at the same time: receivers can still correlate first-arrival times with sender ownership, yielding strong within-round attribution signals.
We therefore add randomized start-time lags.
At the beginning of each round, each client $v$ samples a lag $\ell_v \sim \mathrm{Unif}\{0,1,\ldots,T_{\mathrm{lag}}-1\}$ and delays initiating transmissions by $\ell_v$ slots.
This blurs temporal correlation and increases the likelihood that early transfers are relayed (non-owner) rather than owner chunks. However, time obfuscation can waste uplink capacity during idle waits, motivating tracker-assisted scheduling below.
In our default configuration, $\Delta=1$\,s and $T_{\mathrm{lag}}=3$, i.e., $\ell_v \sim \mathrm{Unif}\{0,1,2\}$ seconds.

\noindent\subsubsection{Tracker-assisted Non-owner-first Scheduling}\label{sec:tracker-scheduling}
Pre-round and time obfuscation improve unlinkability but can add coordination overhead (spray) and/or transient under-utilization (lags).
To reach the target unlinkability level quickly \emph{without} sacrificing throughput, FLTorrent lets the tracker coordinate warm-up transfers. During warm-up, the tracker collects per-client bitfields and assigns transmissions so that, whenever possible, a requested chunk is served by a \emph{non-owner} holder (a client that previously obtained the chunk), rather than by its original source.
This \emph{non-owner-first} preference reduces early owner bias: once a sender has accumulated sufficient non-local chunks, an observer cannot treat ``sender'' as a reliable proxy for ``source.'' Formally, in a warm-up stage starting at slot $s<s_{\mathrm{BT}}$, the tracker knows each client’s inventory $\mathcal{C}_v^r[s]$ and its missing-chunk set
$\mathcal{M}_v^r[s]=\bigl(\bigcup_{u\in\mathcal{N}^r(v)}\mathcal{C}_u^r[s]\bigr)\setminus \mathcal{C}_v^r[s]$.
It schedules feasible missing-chunk transmissions (respecting per-stage capacity and avoiding duplicates) until all \emph{active} clients reach the global cover-set threshold $k_\beta$ (or until $s_{\max}$); at $s_{\mathrm{BT}}$ FLTorrent switches to vanilla BitTorrent.

\textit{Lemma 1 (Hardness of optimal warm-up scheduling).}
Minimizing the warm-up makespan (time until all clients reach the global cover-set threshold $k_\beta$) is NP-complete, by a standard reduction from precedence-constrained makespan scheduling (e.g., $P\mid\mathrm{prec}\mid C_{\max}$)~\cite{Ullman1975,GareyJohnson1979}.

Accordingly, we use tractable heuristics that aim to maximize warm-up bandwidth utilization, which empirically yields low makespan in our settings.

\subsection{Warm-up Scheduling}\label{sec:scheduling}

\begin{figure}[t]
    \centering
    \includegraphics[scale=0.25]{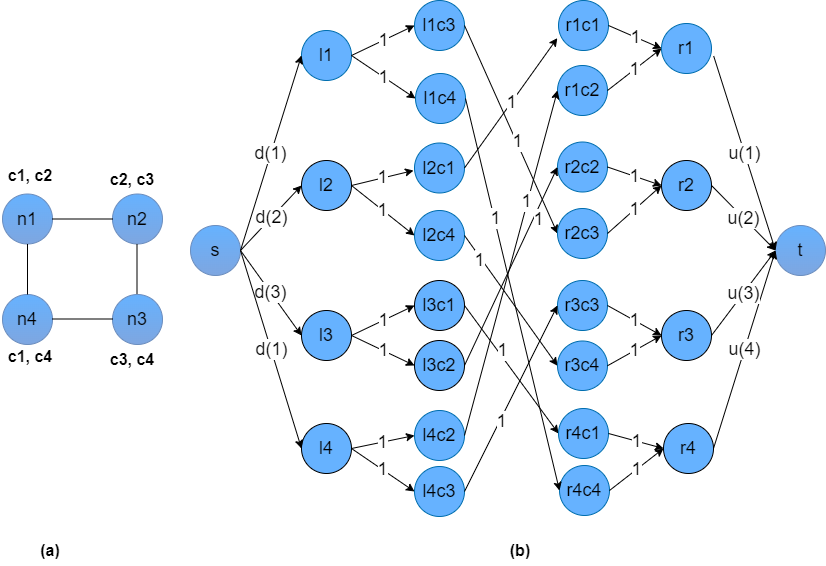}
    \caption{a) Example overlay for a group with 4 nodes and 4 chunks. Each node is depicted with the chunks it holds. b) The resulting max-flow network.}
    \label{fig:MF}
\end{figure}

\noindent\textit{1) Bandwidth-optimal Scheduling as a Max-flow Upper Bound:}
A bandwidth-optimal \emph{stage-wise} warm-up schedule maximizes the group throughput within a stage
(sum of used uplink, equivalently downlink), given current inventories and per-stage chunk budgets.
We schedule in discrete stages of duration $\Delta$.
Each client $v$ can upload at most
$u_v = \lfloor U_v \Delta / C \rfloor$ chunks and download at most
$d_v = \lfloor D_v \Delta / C \rfloor$ chunks per stage.
Stage-wise throughput maximization reduces to a max-flow over a bipartite construction
with supply $u_v$ and demand $d_v$, plus availability edges ensuring a chunk is only sent by
a neighbor that already possesses it.
Figure~\ref{fig:MF} illustrates an example topology and its corresponding max-flow network.
We do not run max-flow online; we use it only as an (offline) upper bound computed with full knowledge of the stage state.

\noindent\textit{2) Practical Schedulers:}
Since max-flow is impractical at scale, we implement centralized and distributed heuristics,
as well as flooding-style dissemination.

\noindent\textit{3) Centralized RandomFIFO:}
The tracker assigns each missing-chunk request to a random client that holds the chunk.
Clients serve assigned chunks FIFO.
A client can send up to $\tau$ chunks simultaneously (BitTorrent commonly uses $\tau=4$).

\noindent\textit{4) Centralized RandomFastestFirst:}
Like RandomFIFO, but a sender prioritizes the $\tau$ fastest requesters based on available
sender uplink and available downlink at the requester.

\noindent\textit{5) Centralized GreedyFastestFirst:}
Like RandomFastestFirst, but the tracker assigns each missing-chunk request to the fastest
feasible sender.

\noindent\textit{6) Distributed Scheduling (Neighborhood-level announcements):}
In distributed scheduling, clients decide which requests to issue and who to serve.
To avoid revealing which specific neighbor holds which chunk during warm-up, the tracker can
provide neighborhood-level availability: for each client $v$, it announces
$\mathcal{C}^{\textrm{TA}}(v,s)=\bigcup_{u\in\mathcal{N}^r(v)}\mathcal{C}_u^r[s]$,
without disclosing the exact mapping chunk $\rightarrow$ neighbor.

\noindent\textit{7) Flooding:}
In flooding, each client sends chunks to neighbors randomly without repetition.
Combined with randomized lags, flooding increases relaying likelihood but wastes bandwidth
and performs worse than coordinated warm-up heuristics.

\noindent\textbf{Non-owner-first refinement.}
When multiple feasible senders can serve a missing chunk, we apply a non-owner-first
tie-breaking preference to the centralized schedulers: the tracker prioritizes a sender that
is not the original source of that chunk; otherwise it falls back to the source.

\subsection{Auditability Under a Deviating Tracker}\label{sec:auditability}

\noindent\textbf{Motivation.}
Although the tracker is not a privacy target (torrent descriptors reveal the chunk-to-owner mapping),
a deviating tracker can still \emph{amplify peer-side attribution} by shaping the environment---e.g.,
generating sparse/imbalanced overlays or issuing warm-up directives that slow early mixing.
This matters in permissioned cross-silo consortia where a biased or faulty operator may
\emph{selectively} weaken some participants' unlinkability, increasing the success probability of
\emph{any} observation-only peer attacker (honest-but-curious neighbors), even without tracker--client collusion.
In permissioned consortia, this mechanism supports external governance; engineering extensions such as replicated trackers or threshold-controlled coordination are complementary and left to future work.

%

%

%

\noindent\textbf{Commit-then-reveal accountability.}
Before seeing per-round inputs, the tracker commits to a public seed $h^r{=}H(\mathsf{seed}^r)$; after the round it reveals $\mathsf{seed}^r$ and a log $\mathsf{Log}^r$ (overlay and warm-up directives).
Clients can recompute the overlay and verify hard constraints (adjacency, configured per-stage caps, and no redundant deliveries except logged retries); otherwise they fail open to vanilla BitTorrent and treat unlinkability guarantees for that round as void.

\subsection{Fault Tolerance in \emph{FLTorrent}}\label{sec:fault-tolerance}
BitTorrent swarming is naturally robust to intermittent connectivity and heterogeneous bandwidth once multiple replicas exist.
Thus, FLTorrent’s key reliability objective is to safely transition through warm-up; after that, BitTorrent provides standard resilience mechanisms (retries, resumption, multi-source downloads, tit-for-tat, dynamic choking).

\noindent\textbf{Within-round dropouts.}
If a client disconnects during a round, the tracker excludes it from \emph{further} scheduling decisions (it no longer issues or serves warm-up assignments).
If the client had already uploaded chunks that have been replicated, dissemination can still complete for remaining participants, and the disconnected client's update may remain in the active set if it is reconstructable by the deadline.
If the disconnected client was the sole holder of some chunks (disconnects before replication), those chunks become unavailable; the round completes over the remaining active set, as in partial participation in practical FL deployments.
Disconnected clients can rejoin in the next round with fresh pseudonyms.

\noindent\textbf{Cross-round churn (joins/leaves).}
Joiners are incorporated at the next round boundary.
Persistent leavers are removed from the next round’s overlay generation.
The tracker regenerates a fresh overlay each round, preventing long-lived neighbor relationships that could amplify cross-round linkage.

%


\noindent\textbf{Byzantine behavior handling.}
We distinguish (a) \emph{integrity-violating} deviations and (b) \emph{liveness-degrading} deviations.
For (a), invalid-chunk injection or payload tampering is detectable via descriptor hash checks~\cite{bep0003} and such content is discarded; under authenticated channels, spoofing another client's round pseudonym is excluded (otherwise it reduces to Sybil/credential compromise as in \S\ref{sec:threat-model}).
For (b), Byzantine peers may lie in bitfields, selectively forward/withhold chunks, send unsolicited chunks, or delay transmissions to slow warm-up.
During warm-up, receivers count only tracker-scheduled triples $(u,w,c)$ toward their cover-set progress; unsolicited transfers are ignored for warm-up accounting and logged as protocol violations.
Repeated violations or lack of progress cause the peer to be marked inactive.
Thus, Byzantine peers can degrade liveness or reveal their own owner chunks, but they cannot force honest senders to violate cover-set gating or owner throttling.
Accordingly, our unlinkability bounds apply to transfers sent by honest senders.
Operationally, we mitigate slowdown by (i) per-peer progress timeouts (clients that fail to make progress are marked inactive), and (ii) evaluating warm-up completion over the remaining active set.
If warm-up cannot complete within $s_{\max}$, the system fails open to vanilla BitTorrent, preserving liveness but voiding unlinkability guarantees for that round (a DoS objective beyond our formal scope).

\noindent\textbf{Privacy implications.}
Because unlinkability guarantees rely on cover-set size and non-owner mass, we evaluate the warm-up threshold over the currently active participants.
A dropout reduces available capacity but does not invalidate unlinkability for other clients, as long as the warm-up threshold is satisfied for the remaining active set.

%% file: Sections/analysis.tex
\section{Unlinkability Analysis and Attacks}
\label{sec:analysis-attacks}

\subsection{Adversary Success Bounds}
\label{sec:probability}

\noindent\textbf{Setting and objective.}
We analyze \emph{within-round source unlinkability} defined in Section~\ref{sec:privacy-metrics} under the threat model in Section~\ref{sec:threat-model}.
Concretely, 
an adversary (possibly a coalition of clients) observes protocol-level signals $\mathcal{O}_v[s]$ (sender round-pseudonyms, piece indices (chunk identifiers), timestamps/order, and warm-up scheduling directives received by $v$) and attempts to attribute an observed transfer to its true source.
We focus on the within-round attribution event $\mathsf{src}(c)=\mathsf{snd}(c)$ for an observed chunk transfer of $c$.
Unless stated otherwise, the bounds below apply to transfers sent by \emph{honest senders} under origin-oblivious chunk selection (e.g., uniform or rarest-first independent of origin), consistent with Adversary~A (and the honest-sender portion of Adversary~B).

\noindent\textbf{Eligible serving buffer and knobs.}
A client always \emph{possesses} its own update chunks locally, but FLTorrent controls which chunks are \emph{eligible for upload} during warm-up.
For a sender $u$ at a serving instant (round $r$, slot $s$), let $\mathcal{B}_u^r[s]$ be the set of currently eligible chunks for upload, and let $B_u := |\mathcal{B}_u^r[s]|$.
The buffer contains (i)~\emph{non-owner} chunks received from other clients and (ii)~a limited number of \emph{owner} chunks from $u$'s update.
Let $O_u$ be the number of eligible owner chunks and $X_u := B_u - O_u$ the eligible non-owner mass at that instant.

\noindent\textbf{Notation clarification.}
Throughout this section, $u$ denotes the \emph{sender} (the client transmitting a chunk) and $v$ denotes the \emph{receiver/observer} (the client receiving the chunk and potentially acting as an attacker), matching $\mathcal{O}_v[s]$.

FLTorrent uses two enforcement knobs during warm-up:
(1)~\textbf{global cover-set gating:} an honest sender enables owner chunks only after its total inventory reaches the global cover threshold
$k_\beta=\lceil \beta|\mathcal{C}^r|\rceil$,
and (2)~\textbf{owner throttling:} at any time $O_u \le \kappa_u$ (default $\kappa_u=1$).
Since client $u$ initially holds its own $K_u^r$ chunks, reaching the global threshold implies that its non-owner inventory is at least
\[
h_u := \max\{0,k_\beta-K_u^r\}.
\]
When serving, honest senders apply an \emph{origin-oblivious} chunk-selection rule: conditioned on $\mathcal{B}_u^r[s]$, the rule does not explicitly privilege owner chunks beyond what is implied by availability/state (e.g., uniform-at-random over $\mathcal{B}_u^r[s]$, or rarest-first restricted to $\mathcal{B}_u^r[s]$).

\noindent\textbf{Proposition 1 (Per-transfer attribution bound).}
Consider a warm-up transfer $u\!\to\!v$ of chunk $c$ sent by an honest sender $u$ under origin-oblivious chunk selection.
If $u$ satisfies global cover-set gating $|\mathcal{C}_u^r[s]|\ge k_\beta$ and owner throttling $O_u\le\kappa_u$, then
\begin{equation}
\Pr\!\big[\mathsf{src}(c){=}u \mid \mathcal{O}_v\big]
\le
\frac{\kappa_u}{\kappa_u+h_u}
=
\frac{\kappa_u}{\kappa_u+\max\{0,k_\beta-K_u^r\}} .
\label{eq:ku_over_k}
\end{equation}

\noindent\emph{Proof.}
Under origin-oblivious selection, the per-transfer attribution posterior equals the eligible owner fraction:
$\Pr[\mathsf{src}(c){=}u \mid \mathcal{O}_v] = O_u/(O_u+X_u)$.
By global cover-set gating, once $u$ is allowed to serve owner chunks, its non-owner mass satisfies
$X_u \ge h_u=\max\{0,k_\beta-K_u^r\}$.
Together with owner throttling ($O_u\le \kappa_u$), Eq.~\eqref{eq:ku_over_k} follows.
With $\kappa_u{=}1$, each warm-up transfer from an honest sender is bounded by $1/(1+h_u)$.
Thus, increasing the global cover ratio $\beta$ enlarges the required non-owner mass and decreases the attribution posterior.
For intuition, a neighborhood-level ``random guessing'' baseline corresponds to posteriors on the order of $1/|\mathcal{N}^r(v)|$ for an observer $v$; our goal is to drive per-transfer attribution toward this regime while keeping warm-up short.

\smallskip
\noindent\textbf{Warm-up mixing further tightens the posterior (high probability).}
Warm-up increases the eligible non-owner mass $X_u$, making $O_u/B_u$ typically smaller than the deterministic global-threshold cap $\kappa_u/(\kappa_u+h_u)$.
Write $X_u = Z_R(u) + Z_T(u)$, where $Z_R(u)$ is injected by pre-round spray and $Z_T(u)$ is gained early via randomized lags and non-owner-first scheduling.

\smallskip
\noindent\emph{Pre-round obfuscation.}
Each sender sprays $\sigma=\lfloor R\cdot K_u^r \rfloor$ chunks to uniformly random non-neighbors.
For a fixed $u$, let $\mu_u:=\mathbb{E}[Z_R(u)]$; under random assignment,
\[
\textstyle
\mu_u=\sum_{v:\,u\notin\mathcal{N}^r(v)\cup\{v\}}\tfrac{\sigma}{n-1-|\mathcal{N}^r(v)|},
\]
so $\mu_u=\sigma$ for an $m$-regular overlay and $\mu_u\approx\sigma$ for near-regular overlays.
Since $Z_R(u)$ is a Poisson-binomial sum, a Chernoff-type bound yields
$\Pr[Z_R(u)\le (1{-}\epsilon)\mu_u]\le \exp(-\epsilon^2\mu_u/2)$ for $\epsilon\in(0,1)$.

\smallskip
\noindent\emph{Time obfuscation and non-owner-first scheduling.}
If lags are i.i.d.\ uniform on $\{0,\ldots,T_{\mathrm{lag}}{-}1\}$, then
$p_{\text{lead}}=\Pr[\ell_v<\ell_u]=(T_{\mathrm{lag}}{-}1)/(2T_{\mathrm{lag}})$.
With average degree $m$ and availability factor $q\in(0,1]$, we have
$\mathbb{E}[Z_T(u)]\ge m\,p_{\text{lead}}\,q$ and
$\Pr[Z_T(u)\le (1{-}\epsilon)\mathbb{E}[Z_T(u)]]\le \exp(-\epsilon^2\mathbb{E}[Z_T(u)]/2)$.
In practice, $q$ can be estimated online from warm-up bitfields as the empirical fraction of neighbor-request pairs for which at least one non-owner holder is available; setting $q=0$ gives a conservative fallback to the deterministic gating-only bound in Eq.~\eqref{eq:ku_over_k}.

\smallskip
\noindent
By a union bound, with probability at least $1-\eta$ where
$\eta=\exp(-\epsilon^2\mu_u/2)+\exp(-\epsilon^2\mathbb{E}[Z_T(u)]/2)$, we have
$X_u\ge (1{-}\epsilon)\big(\mu_u + m\frac{T_{\mathrm{lag}}-1}{2T_{\mathrm{lag}}}q\big)$, hence
\begin{equation}
\tfrac{O_u}{B_u}
=
\tfrac{O_u}{O_u+X_u}
\le
\min\!\left\{
\tfrac{\kappa_u}{\kappa_u+h_u},
\tfrac{\kappa_u}{\kappa_u+(1{-}\epsilon)\big(\mu_u + m\tfrac{T_{\mathrm{lag}}-1}{2T_{\mathrm{lag}}}q\big)}
\right\}.
\label{eq:mixing_bound}
\end{equation}
Thus, the deterministic global-threshold cap in Eq.~\eqref{eq:ku_over_k} always applies, while pre-round spray, randomized lags, and non-owner-first scheduling typically tighten it further by increasing the realized non-owner mass.


\noindent\textbf{Repeated observations from the same sender.}
If an attacker observes $s_u$ transfers from the same honest sender $u$ during warm-up, with selection without replacement, and if
$B_\star=\min_t B_u(t)$ and $O_\star=\max_t O_u(t)$ across those $s_u$ draws, then
\begin{multline}
\Pr\big[\exists\text{ owner chunk in } s_u \text{ transfers}\big]
= 1 - \frac{\binom{B_\star-O_\star}{s_u}}{\binom{B_\star}{s_u}}\\
\le
1-\Big(1-\frac{O_\star}{B_\star-s_u+1}\Big)^{s_u}
\le
\min\!\left\{1,\,
s_u\cdot\frac{\kappa_u}{\kappa_u+h_u}
\right\}.
\label{eq:multi}
\end{multline}
This bound explains why empirical ASR can exceed the single-transfer posterior when attackers aggregate evidence across multiple transfers, even though each individual transfer remains capped.

\noindent\textbf{Conditionality under an auditable tracker.}
Under the \emph{auditable tracker} model in \S\ref{sec:threat-model}, the bounds above are conditional on the published round log passing verification of the verifiable constraints in \S\ref{sec:auditability}.
If violations are detected, clients fail open to vanilla BitTorrent and treat unlinkability guarantees for that round as void.

\subsection{Collusion-aware Bounds (Alliance Filtering)}\label{sec:collusion}
We incorporate a strong coalition that can recognize and discount some non-owner traffic originating from coalition members.
Fix an honest sender $u$ at a serving instant.
Let $\rho_u\in[0,1]$ denote the fraction of $X_u$ originating from the coalition, and let $\phi\in[0,1]$ be the coalition's \emph{recognition strength}
($\phi{=}0$: no filtering; $\phi{=}1$: perfect erasure).
We treat $\phi$ as an analysis parameter capturing coalition side information; in evaluation we report envelopes spanning $\phi\in\{0,1\}$ to bracket practical recognition strength.
Alliance filtering reduces effective non-owner mass to $X_u^{\mathrm{eff}}=(1{-}\phi\rho_u)X_u$, giving
\begin{equation}
\theta_u^{\mathrm{AF}}
=
\tfrac{O_u}{O_u + X_u^{\mathrm{eff}}}
\le
\min\!\Big\{
\tfrac{\kappa_u}{\kappa_u+h_u},\;
\tfrac{\kappa_u}{\kappa_u + (1{-}\phi\rho_u)X_u}
\Big\}.
\label{eq:af_bound}
\end{equation}
Coalition filtering loosens the realized mixing term by $(1{-}\phi\rho_u)$ but cannot beat the global-threshold cap $\kappa_u/(\kappa_u+h_u)$ for honest senders.

\noindent\textbf{High-probability mixing under collusion.}
Let $X_u = Z_R(u)+Z_T(u)$ as before. With probability at least $1{-}\eta$,
\begin{equation}
\theta_u^{\mathrm{AF}}
\le
\min\!\Big\{
\tfrac{\kappa_u}{\kappa_u+h_u},\,
\tfrac{\kappa_u}{\kappa_u + (1{-}\phi\rho_u)(1{-}\epsilon)\big(\mu_u {+} m\tfrac{T_{\mathrm{lag}}-1}{2T_{\mathrm{lag}}}q\big)}
\Big\}.
\label{eq:af_mixing}
\end{equation}
These reduce to the non-colluding bounds when $\phi{=}0$ and give the ideal-collusion envelope when $\phi{=}1$.


\noindent\textbf{Repeated observations with collusion.}
For $s_u$ observations from the same sender, Eq.~\eqref{eq:multi} extends to
\begin{equation}
\Pr\!\big[\exists\text{ owner in } s_u \text{ draws}\big]
\le
s_u
\min\!\Big\{
\tfrac{\kappa_u}{\kappa_u+h_u},\,
\tfrac{\kappa_u}{\kappa_u + (1{-}\phi\rho_u)X_u}
\Big\}.
\label{eq:af_multi}
\end{equation}
Collusion increases the effective observation count and reduces obfuscating mass, but the global-threshold cap $\kappa_u/(\kappa_u+h_u)$ continues to govern attribution success for honest senders (the bound is a union bound and may be loose for large $s_u$).

\subsection{Attack Strategies and Metrics}
\label{attack}

\noindent\textbf{Why we focus on attribution attacks.}
Standard defenses against membership inference or reconstruction attacks~\cite{shokri2017membership,zhu2019deep,geiping2020inverting}
are orthogonal and can be layered atop FLTorrent.
Our focus is solely on \emph{within-round attribution} introduced by BitTorrent-style dissemination, i.e., inferring whether an observed sender is also the true source of the transmitted chunk/update.

\noindent\textbf{Metric: Attribution Success Rate (ASR).}
For a receiver $v$, consider all warm-up transfers it observes in a round.
An attacker outputs, for each observed transfer, a binary attribution decision (or a ranked guess) for whether $\mathsf{src}(c)=\mathsf{snd}(c)$.
We report ASR as the fraction of observed transfers that are attributed correctly; in evaluation we often use the \emph{maximum} ASR over receivers (and over coalition members) as a conservative summary.
This metric is \emph{empirically} related to the bounds in Section~\ref{sec:probability}: Eq.~\eqref{eq:ku_over_k} caps per-transfer attribution for honest senders, while Eq.~\eqref{eq:multi} captures how aggregating evidence across multiple transfers can increase success.

\noindent\textbf{Three observation-only attack strategies.}
All attacks below are \emph{observation-only} and fit Adversary~A in Section~\ref{sec:threat-model} (honest-but-curious, possibly colluding). They do not inspect payload contents or manipulate protocol state.

\noindent(1) \textbf{Sequential Greedy Strategy.}
The attacker labels the \emph{first} chunk received from each sender pseudonym as its owner, targeting the strongest early-round signal and instantiating the ``early owner bias'' that warm-up aims to suppress.

\noindent(2) \textbf{Amount Greedy Strategy.}
For each observed sender pseudonym, the attacker aggregates evidence across transfers by grouping received pieces by their update/torrent descriptor (the adversary does not know the producing client's real identity, only descriptor identities).
It then attributes the sender to the descriptor that appears most frequently among the sender's early transfers.
Intuitively, if a sender transmits disproportionately many chunks from its own (unknown-to-peers) update early, frequency becomes a strong attribution cue.

\noindent(3) \textbf{Clustering Strategy.}
The attacker forms a feature vector per sender pseudonym using (i) temporal features (arrival order statistics) and (ii) frequency features (counts per observed descriptor/chunk family), then clusters sender pseudonyms against descriptor-level IDs and outputs the best match.
This combines the early-time and volume signals and captures more sophisticated inference without requiring payload inspection.

\noindent\textbf{Relation to our bounds.}
Sequential Greedy approximates the strongest single-shot posterior, while Amount Greedy and Clustering exploit multiple observations ($s_u$) from a sender.
Equations~\eqref{eq:ku_over_k} and~\eqref{eq:multi} therefore upper-bound attacker advantage as the global cover threshold $k_\beta$ grows through the induced non-owner mass $h_u=\max\{0,k_\beta-K_u^r\}$; the bound tightens further with realized mixing in Eq.~\eqref{eq:mixing_bound}, matching the empirical ASR trends reported in our evaluation.

%% file: Sections/6experiment.tex
\usepgfplotslibrary{fillbetween}
\section{Performance Evaluation}

This section evaluates \emph{FLTorrent} along four axes:
(i) \textbf{learning utility} (accuracy vs.\ CFL and GossipDFL),
(ii) \textbf{warm-up efficiency} (bandwidth utilization vs.\ a max-flow upper bound and heuristics),
(iii) \textbf{end-to-end round cost} (warm-up share and scalability), and
(iv) \textbf{privacy efficacy} under \emph{within-round source unlinkability} objective (\textbf{attack success rate}, ASR).

%

\noindent\textbf{Semantics and scope of evaluation.}
FLTorrent follows the FedAvg update rule and differs from CFL only in the \emph{dissemination substrate}.
As in Section~\ref{sec:system-model}, each client aggregates over the \emph{reconstructable active set} by the round deadline.
In our learning experiments, we set the deadline so updates are reconstructable with high probability; otherwise, we fall back to standard partial-participation semantics (FedAvg over the reconstructable active set) under stragglers or disconnects.

\noindent\textbf{Metrics.}
We report:
(1) \textbf{Bandwidth utilization} during warm-up;
(2) \textbf{Warm-up duration} $T_{\text{warm}}$ (until \emph{active} clients reach the warm-up threshold and switch to vanilla BitTorrent);
(3) \textbf{Round duration} $T_{\text{round}}$ (warm-up + BitTorrent dissemination; local aggregation is executed after dissemination);
(4) \textbf{ASR}: the fraction of correct attributions under the attack models in Section~\ref{attack} and threat model in Section~\ref{sec:threat-model}.

\subsection{Setting}

\noindent\textbf{Model and chunking.}
Unless otherwise stated, we slice a \emph{GoogLeNet} update into $206$ chunks of $256\,\mathrm{KiB}$ each ($206\times256\,\mathrm{KiB}=51.5\,\mathrm{MiB}\approx 54.0\,\mathrm{MB}$ in decimal units, consistent with the $\sim\!51$--$54\,\mathrm{MB}$ scale reported in~\cite{shu2018if}).

\noindent\textbf{Access links.}
We sample heterogeneous uplink/downlink capacities based on European residential broadband statistics~\cite{OECD2024}: uplink $15.5$--$25.3\,\mathrm{Mbps}$ and downlink $36.5$--$121\,\mathrm{Mbps}$, corresponding to roughly $[7,12]$ and $[18,60]$ chunks/s respectively for $256\,\mathrm{KiB}$ chunks.

\noindent\textbf{Overlay.}
We use a random overlay with minimum degree $m$ (default $m{=}10$) and heterogeneous neighbor counts above $m$.

\noindent\textbf{Warm-up knobs (unlinkability hardening).}
We evaluate $\beta$ (global cover-set ratio), \texttt{PR} (pre-round obfuscation), and \texttt{TL} (time obfuscation).
By default, $R{=}0.2$ for \texttt{PR} ($41$ chunks per update), $\ell_v \sim \mathrm{Unif}\{0,1,2\}$ seconds, and $\beta=10\%$.
Sweeping $\beta$ means setting $k_\beta=\lceil \beta|\mathcal{C}^r|\rceil$, where $|\mathcal{C}^r|=\sum_{v\in\mathcal{V}}K_v^r$; for $n{=}500$ and $\beta=10\%$, this gives $k_\beta=10{,}300$ pieces ($\approx 2.5$\,GiB per client at $256$\,KiB/piece).
This default is a privacy-oriented stress configuration; smaller $\beta$ reduces storage and warm-up cost.

\subsection{FLTorrent vs.\ Gossip DFL vs.\ CFL}

We compare \emph{FLTorrent} with CFL and \emph{Gossip DFL}~\cite{lian2017can,koloskova2019decentralized} on \textsc{MNIST} and \textsc{CIFAR-10} under IID and Dirichlet Non-IID splits ($\alpha\in\{0.1,0.5,1.0\}$).
Each run lasts 50 communication rounds; Table~\ref{tab:accuracy} reports mean test accuracy over 10 seeds, using identical hyperparameters across methods.

\begin{table}[t]
\centering
\caption{Test accuracy (mean$\pm$std) on CIFAR-10 and MNIST ($n{=}50$ clients, local epochs${}=5$, batch size${}=32$, 50 communication rounds). FLTorrent consistently surpasses GossipDFL and remains close to CFL.}
\label{tab:accuracy}
\begin{tabular*}{\columnwidth}{@{\extracolsep{\fill}}llccc@{}}
\toprule
& Distribution & CFL & GossipDFL & FLTorrent \\
\midrule
\multirow{4}{*}{\rotatebox{90}{CIFAR-10}}
  & Dir($\alpha$=0.1) & 0.624$\pm$0.025 & 0.379$\pm$0.031 & 0.621$\pm$0.027 \\
  & Dir($\alpha$=0.5) & 0.667$\pm$0.023 & 0.633$\pm$0.027 & 0.671$\pm$0.024 \\
  & Dir($\alpha$=1.0) & 0.724$\pm$0.023 & 0.709$\pm$0.025 & 0.722$\pm$0.022 \\
  & IID               & 0.793$\pm$0.017 & 0.734$\pm$0.021 & 0.791$\pm$0.019 \\
\midrule
\multirow{4}{*}{\rotatebox{90}{MNIST}}
  & Dir($\alpha$=0.1) & 0.979$\pm$0.006 & 0.907$\pm$0.007 & 0.978$\pm$0.007 \\
  & Dir($\alpha$=0.5) & 0.987$\pm$0.005 & 0.965$\pm$0.004 & 0.988$\pm$0.004 \\
  & Dir($\alpha$=1.0) & 0.989$\pm$0.003 & 0.972$\pm$0.004 & 0.990$\pm$0.003 \\
  & IID               & 0.992$\pm$0.002 & 0.986$\pm$0.003 & 0.992$\pm$0.002 \\
\bottomrule
\end{tabular*}
\end{table}
\vspace{-4pt}
\vspace{4pt}
Overall, FLTorrent matches (closely tracks) CFL and consistently outperforms GossipDFL, especially under strong heterogeneity where local mixing attenuates global information.

\subsection{End-to-End Round Communication Cost}
\label{sec:cost_decomposition}

\begin{figure}[t]
    \centering
    \input{tex/bandwidth_timeslots_seconds.tex}
    \vspace{-10pt}
    \caption{Warm-up bandwidth utilization for online heuristics vs.\ a max-flow upper bound.}
    \label{fig:bandwidth}
\end{figure}
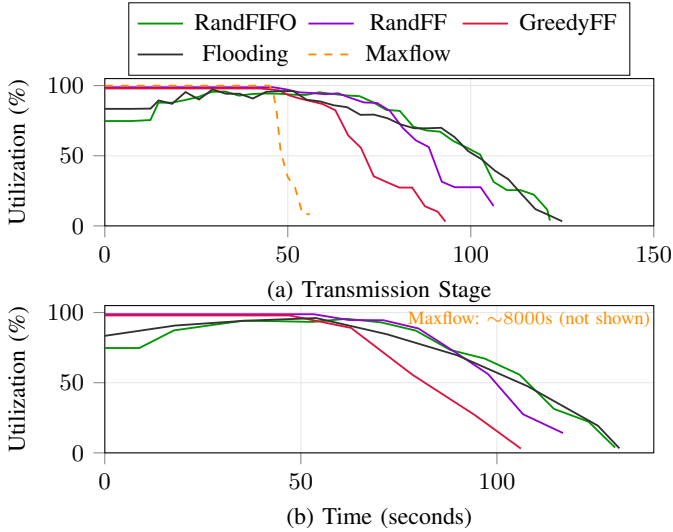

We next quantify the communication overhead of unlinkability hardening, focusing on
(i) warm-up bandwidth efficiency and (ii) end-to-end round time decomposition.
Figure~\ref{fig:bandwidth} shows that warm-up sustains high utilization across heuristics.
\textbf{GreedyFastestFirst} completes warm-up earlier while approaching the max-flow upper bound in the high-utilization regime; unless otherwise stated, we use it below.
\vspace{-5pt}
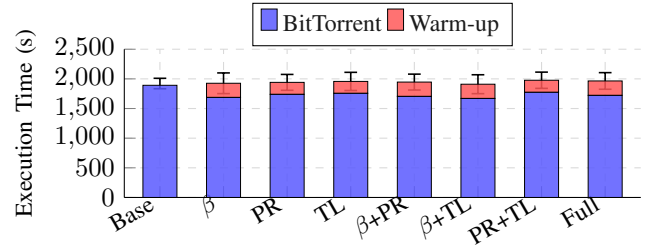
\begin{figure}[t]
    \centering
    \input{tex/privacy_ablation}
    \caption{End-to-end time decomposition under privacy ablations ($\beta$, \texttt{PR}, \texttt{TL}), split into the BitTorrent and warm-up phases.}
    \label{fig:normalized_warmup_fixed}
\end{figure}

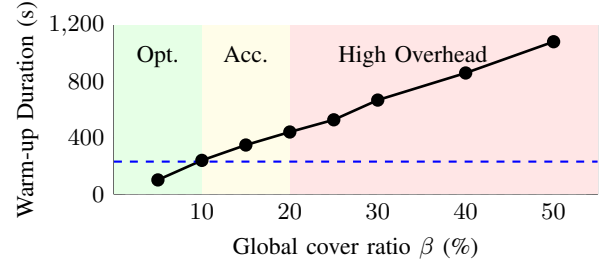
\begin{figure}[t]
    \centering
    \input{tex/k_warmup}
    \caption{Warm-up duration as the global cover ratio $\beta$ increases, where $k_\beta=\lceil \beta|\mathcal C^r|\rceil$.}
    \label{fig:k_sweep_dual}
\end{figure}

\vspace{4pt}
\noindent\textbf{Control plane.}
The tracker processes $O(nm)$ neighborhood-availability entries per stage and emits at most $O(\sum_v d_v)$ scheduled transfers per stage.
In our 100--500 peer experiments, bitfields and stage schedules are only a few KiB per client and are negligible relative to the multi-GiB data plane; tracker integrity is discussed in \S\ref{sec:auditability}.
Figure~\ref{fig:normalized_warmup_fixed} decomposes the round time (100 nodes).
With all defenses enabled (\texttt{Full}), warm-up takes $243.32$\,s and the subsequent BitTorrent phase takes $1721.75$\,s (total $1965.07$\,s).
Compared to BitTorrent-only (\texttt{Base}, $1891.75$\,s), the \emph{total} overhead is modest ($\approx 3.9\%$), because warm-up improves early chunk diversity and shortens the later BitTorrent phase.
Within \texttt{Full}, adding \texttt{PR} and \texttt{TL} on top of the global cover ratio $\beta$ has small marginal cost (warm-up $238.78$\,s $\rightarrow$ $243.32$\,s).
Figure~\ref{fig:k_sweep_dual} quantifies the deployment-time privacy--cost knob $\beta$.
Warm-up ends once every \emph{active} client reaches the global cover threshold $k_\beta$ (\S\ref{sec:warmup}).
As $\beta$ increases, warm-up grows monotonically: $\approx 99.5$\,s at $\beta=5\%$, $\approx 238.8$\,s at $\beta=10\%$, and $\approx 1084.7$\,s at $\beta=50\%$.
Thus, moderate thresholds provide substantial early-round mixing at manageable cost, whereas aggressive thresholds can dominate end-to-end time.
\vspace{-5pt}
\begin{table}[t]
\centering
\caption{End-to-end round cost under \texttt{Full} privacy (GreedyFastestFirst, 51\,MB model, chunk=256\,KiB).}
\label{tab:privacy_cost}
\begin{tabular*}{\columnwidth}{@{\extracolsep{\fill}}rcccc@{}}
\toprule
$n$ & $T_{\text{warm}}$ (s) & Share (\%) & Util. (\%) & $T_{\text{round}}$ (s) \\
\midrule
100 & 243 & 12.4 & 80.5 & 1965 \\
200 & 472 & 11.5 & 76.3 & 4118 \\
300 & 730 & 11.7 & 77.1 & 6229 \\
400 & 939 & 11.5 & 79.4 & 8192 \\
500 & 1229 & 11.7 & 75.4 & 10501 \\
\bottomrule
\end{tabular*}
\end{table}
\vspace{4pt}

Table~\ref{tab:privacy_cost} shows stable scaling from 100 to 500 peers:
warm-up remains a consistent $\approx 11.5\%$--$12.4\%$ share of round time and utilization stays in the $75\%$--$80\%$ range.

\subsection{Privacy Evaluation: Within-round Source Unlinkability}

We now quantify privacy under the threat model in Section~\ref{sec:threat-model} using the attacks in Section~\ref{attack}.
Unless otherwise stated, we use a 100-node overlay with minimum degree $m{=}10$.
As a coarse reference point, if an adversary cannot do better than \emph{uniform guessing among $m$ candidate neighbors} for first-hop attribution, its success rate is $\approx 1/m$ (10\% at $m{=}10$).
Our goal is to push empirical ASR toward this regime.

\begin{figure}[t]
    \centering
    \input{tex/privacy_ablation_asr}
    \caption{Privacy ablation: maximum ASR under three inference strategies (Sequence, Count, Cluster).}
    \label{fig:privacy_asr_ablation}
\end{figure}
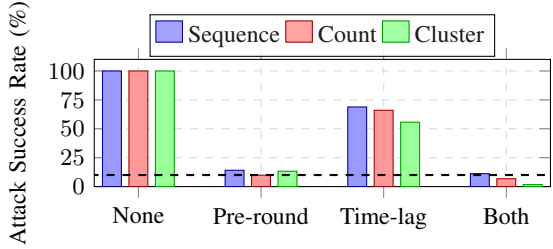

\input{tex/4_privacy_plots}

\noindent\textbf{Connecting ASR to the analysis.}
Section~\ref{sec:probability} bounds the \emph{per-transfer} attribution posterior for honest senders (Eq.~\eqref{eq:ku_over_k}) and its warm-up tightening (Eq.~\eqref{eq:mixing_bound}), whereas ASR aggregates \emph{multiple} observations from a sender (cf.\ Eq.~\eqref{eq:multi}); hence empirical ASR need not match the single-transfer cap.
Figure~\ref{fig:privacy_asr_ablation} shows that hardening is necessary: without defenses ASR is near-perfect; pre-round obfuscation gives the largest drop; time-lags alone are insufficient; and combining them approaches the neighborhood random-guess baseline.
Figure~\ref{fig:privacy_attack_analysis} shows that overlay density is the strongest lever: increasing $m$ from 5 to 25 reduces max ASR from 26.99\% to 4.29\%, with Figure~\ref{fig:privacy_attack_analysis}(a) plotting the moving $\approx 1/m$ baseline.
Pre-round volume has diminishing returns: increasing $R$ from 10\% to 50\% changes max ASR only from 11.43\% to 11.27\%.
Larger networks further amplify privacy: as $n$ grows from 100 to 500, Sequence ASR drops from 10.90\% to 7.31\%, while Count ASR rises slightly from 5.58\% to 6.78\%, still below $1/m$; with fixed $m=10$, all attacks remain close to or below random guessing, consistent with more relaying and larger non-owner cover mass.
Finally, under collusion, the chance that at least one attacker succeeds rises from 13.56\% ($a{=}5$) to 30.82\% ($a{=}25$), but per-attacker ASR stays low, 11.31\%--14.32\%, indicating resilience to observation-only collusion.

\subsection{LLM-Scale Round-Time Overhead}

\begin{figure}[t]
    \centering
    \input{tex/llm_time}
    \caption{Round duration overhead of FLTorrent with unlinkability hardening vs.\ BitTorrent-only for LLM-scale updates.}
    \label{fig:time_llm}
\end{figure}

We use LLM-scale runs only as dissemination stress tests, not as learning-convergence evidence.
Figure~\ref{fig:time_llm} compares FLTorrent with BitTorrent-only over 7--10\,Gbps links for four large update artifacts corresponding to Gemma-7B, DeepSeek-R1-14B, Qwen2.5-32B, and Llama-3.3-70B.
FLTorrent increases round time by 9.97\%, 6.60\%, 7.09\%, and 10.01\%, respectively, using the same warm-up mechanisms evaluated above.

%% file: tex/bandwidth_timeslots_seconds.tex
\definecolor{mygreen}{RGB}{0,150,0}
\definecolor{mypurple}{RGB}{148,0,211}
\definecolor{myred}{RGB}{220,20,60}
\definecolor{myblack}{RGB}{50,50,50}
\definecolor{myorange}{RGB}{255,140,0}
\definecolor{gridcolor}{RGB}{230,230,230}

\begin{tikzpicture}
\begin{groupplot}[
    group style={
        group size=1 by 2,          
        vertical sep=30pt,          
        xlabels at=edge bottom,
    },
    width=0.82\columnwidth,         
    height=0.22\columnwidth,        
    scale only axis,
    grid=major,
    major grid style={line width=0.2pt, color=gridcolor},
    tick align=outside,
    tick pos=left,
    every axis/.append style={font=\small},
]

\nextgroupplot[
    xlabel={(a) Transmission Stage},
    ylabel={Utilization (\%)},
    xmin=0, xmax=150,
    ymin=0, ymax=105,
    xtick={0,50,100,150},
    ytick={0,50,100},
    legend style={
        draw=black,
        line width=0.3pt,
        fill=white,
        legend columns=3,
        at={(0.5,1.02)},
        anchor=south,
        column sep=2pt,
        font=\small,
    },
]

\addplot[color=mygreen, line width=0.7pt, mark=none] coordinates {
    (0,74.75) (7.33,74.75) (12.46,75.44) (14.65,87.84) (18.32,87.84)
    (21.98,89.73) (25.64,91.97) (29.31,95.59) (32.97,95.76) (36.64,92.83)
    (40.30,94.04) (43.96,94.38) (47.63,94.21) (51.29,93.52) (54.95,93.35)
    (58.62,95.25) (62.28,94.04) (65.94,93.01) (69.61,92.49) (73.27,88.18)
    (76.93,82.67) (80.60,81.98) (84.26,70.79) (87.92,68.03) (91.59,67.17)
    (95.25,60.45) (98.91,55.80) (102.58,50.81) (106.24,31.35) (109.90,25.49)
    (113.57,25.49) (117.23,22.22) (120.89,11.54) (121.63,3.79)
};
\addlegendentry{RandFIFO}

\addplot[color=mypurple, line width=0.7pt, mark=none] coordinates {
    (0,98.84) (7.08,98.84) (14.16,98.84) (21.25,98.84) (28.33,98.84)
    (35.41,98.84) (42.49,98.84) (46.03,98.84) (49.57,97.29) (53.12,95.23)
    (56.66,94.71) (60.20,93.85) (63.74,94.54) (67.28,91.09) (70.82,87.99)
    (74.36,87.65) (77.90,81.97) (81.44,69.74) (84.99,60.96) (88.53,56.14)
    (92.07,31.51) (95.61,27.55) (99.15,27.55) (102.69,27.55) (106.24,14.12)
};
\addlegendentry{RandFF}

\addplot[color=myred, line width=0.7pt, mark=none] coordinates {
    (0,97.97) (7.00,97.97) (14.00,97.97) (21.00,97.97) (28.00,97.97)
    (35.00,97.97) (42.00,97.97) (44.80,97.46) (47.59,96.61) (50.39,93.02)
    (53.19,90.80) (55.98,89.10) (59.48,86.88) (62.99,82.44) (66.49,64.52)
    (69.99,55.64) (73.49,35.33) (76.99,31.24) (80.48,27.31) (83.98,27.31)
    (87.48,13.99) (90.98,10.07) (93.07,3.07)
};
\addlegendentry{GreedyFF}

\addplot[color=myblack, line width=0.7pt, mark=none] coordinates {
    (0,83.43) (7.35,83.43) (12.50,83.61) (14.70,89.42) (18.38,87.03)
    (22.06,95.40) (25.73,90.10) (29.41,97.28) (33.08,94.21) (36.76,94.21)
    (40.44,90.79) (44.11,95.57) (47.79,96.09) (51.46,96.09) (55.14,89.93)
    (58.82,88.73) (62.49,85.99) (66.17,84.63) (69.84,79.16) (73.52,79.33)
    (77.20,76.77) (80.87,72.32) (84.55,69.59) (88.22,69.59) (91.90,69.93)
    (95.58,63.26) (99.25,53.34) (102.93,47.53) (106.60,39.32) (110.28,33.17)
    (113.96,22.40) (117.63,11.97) (121.31,7.69) (124.98,3.25)
};
\addlegendentry{Flooding}

\addplot[color=myorange, line width=0.7pt, mark=none, dashed] coordinates {
    (0,100) (10,100) (20,100) (30,100) (40,100) (45,100)
    (46,95.48) (47,80.65) (48,54.19) (49,43.87) (50,34.84)
    (51,30.97) (52,27.10) (53,18.06) (54,8.39) (55,8.39) (56,8.39)
};
\addlegendentry{Maxflow}

\nextgroupplot[
    xlabel={(b) Time (seconds)},
    ylabel={Utilization (\%)},
    xmin=0, xmax=140,
    ymin=0, ymax=105,
    xtick={0,50,100},
    ytick={0,50,100},
    clip=false,
]

\addplot[color=mygreen, line width=0.7pt, mark=none] coordinates {
    (0,74.75) (8.81,74.75) (17.63,87.32) (26.44,90.77) (35.24,94.21)
    (44.05,93.87) (52.87,93.35) (61.68,95.59) (70.49,93.01) (79.31,87.15)
    (88.12,73.37) (96.92,67.17) (105.74,55.80) (114.55,31.35) (123.36,22.22)
    (130.14,3.79)
};

\addplot[color=mypurple, line width=0.7pt, mark=none] coordinates {
    (0,98.84) (17.77,98.84) (35.54,98.84) (53.32,98.84) (62.21,94.71)
    (71.09,94.54) (79.98,88.68) (88.87,72.67) (97.75,56.14) (106.64,27.55)
    (116.86,14.12)
};

\addplot[color=myred, line width=0.7pt, mark=none] coordinates {
    (0,97.97) (31.39,97.97) (47.09,97.97) (62.78,89.10) (78.48,55.64)
    (94.17,27.31) (106.10,3.07)
};

\addplot[color=myblack, line width=0.7pt, mark=none] coordinates {
    (0,83.43) (17.96,90.79) (35.93,94.21) (53.89,96.09) (71.86,84.63)
    (89.83,69.59) (107.80,47.53) (125.76,19.32) (131.23,3.25)
};

\node[anchor=west, font=\scriptsize, color=myorange] at (axis cs:75,95) {Maxflow: ${\sim}8000$s (not shown)};

\end{groupplot}
\end{tikzpicture}

%% file: tex/privacy_ablation.tex
\begin{tikzpicture}
\begin{axis}[
    ybar stacked,
    width=0.95\columnwidth,
    height=0.4\columnwidth,
    bar width=0.45cm,
    ylabel={Execution Time (s)},
    ylabel style={font=\small},
    axis x line*=bottom,
    axis y line*=left,
    xtick=data,
    xticklabels={Base, $\beta$, PR, TL, $\beta$+PR, $\beta$+TL, PR+TL, Full},
    x tick label style={
        rotate=30,
        anchor=east,
        font=\small
    },
    ytick={0,500,1000,1500,2000,2500},
    ymin=0, ymax=2500,
    grid=major,
    grid style={dashed,gray!30},
    enlarge x limits=0.08,
    every axis plot/.append style={
        fill opacity=0.85,
        draw=black,
        line width=0.3pt
    },
    legend style={
        at={(0.5,1.02)},
        anchor=south,
        legend columns=-1,
        font=\small,
        fill=white,
        draw=black,
        fill opacity=0.9
    },
    legend cell align={left}
]

\addplot[fill=blue!65] coordinates {
    (0, 1891.75)  (1, 1687.31)  (2, 1738.96)  (3, 1756.18)
    (4, 1704.53)  (5, 1670.09)  (6, 1773.40)  (7, 1721.75)
};
\addlegendentry{BitTorrent};

\addplot[fill=red!65] coordinates {
    (0, 0)       (1, 238.78)  (2, 202.22)  (3, 201)
    (4, 241.5)   (5, 239.92)  (6, 203.5)   (7, 243.32)
};
\addlegendentry{Warm-up};

\draw[black, line width=0.6pt] 
    (axis cs:0,2009.84) -- (axis cs:0,1833.66)
    (axis cs:-0.1,2009.84) -- (axis cs:0.1,2009.84)
    (axis cs:-0.1,1833.66) -- (axis cs:0.1,1833.66);
\draw[black, line width=0.6pt]
    (axis cs:1,2100.45) -- (axis cs:1,1751.73)
    (axis cs:0.9,2100.45) -- (axis cs:1.1,2100.45)
    (axis cs:0.9,1751.73) -- (axis cs:1.1,1751.73);
\draw[black, line width=0.6pt]
    (axis cs:2,2075.06) -- (axis cs:2,1807.30)
    (axis cs:1.9,2075.06) -- (axis cs:2.1,2075.06)
    (axis cs:1.9,1807.30) -- (axis cs:2.1,1807.30);
\draw[black, line width=0.6pt]
    (axis cs:3,2110.36) -- (axis cs:3,1804.00)
    (axis cs:2.9,2110.36) -- (axis cs:3.1,2110.36)
    (axis cs:2.9,1804.00) -- (axis cs:3.1,1804.00);
\draw[black, line width=0.6pt]
    (axis cs:4,2080.07) -- (axis cs:4,1811.99)
    (axis cs:3.9,2080.07) -- (axis cs:4.1,2080.07)
    (axis cs:3.9,1811.99) -- (axis cs:4.1,1811.99);
\draw[black, line width=0.6pt]
    (axis cs:5,2069.49) -- (axis cs:5,1750.53)
    (axis cs:4.9,2069.49) -- (axis cs:5.1,2069.49)
    (axis cs:4.9,1750.53) -- (axis cs:5.1,1750.53);
\draw[black, line width=0.6pt]
    (axis cs:6,2113.94) -- (axis cs:6,1839.86)
    (axis cs:5.9,2113.94) -- (axis cs:6.1,2113.94)
    (axis cs:5.9,1839.86) -- (axis cs:6.1,1839.86);
\draw[black, line width=0.6pt]
    (axis cs:7,2105.17) -- (axis cs:7,1824.97)
    (axis cs:6.9,2105.17) -- (axis cs:7.1,2105.17)
    (axis cs:6.9,1824.97) -- (axis cs:7.1,1824.97);

\end{axis}
\end{tikzpicture}

%% file: tex/k_warmup.tex
\begin{tikzpicture}
\begin{axis}[
    width=0.44\textwidth,
    height=0.21\textwidth,
    xlabel={Global cover ratio $\beta$ (\%)},
    ylabel={Warm-up Duration (s)},
    ylabel style={yshift=-3pt},
    xmin=0, xmax=55,
    ymin=0, ymax=1200,
    xtick={10,20,30,40,50},
    ytick={0,400,800,1200},
    grid=major,
    grid style={dashed, gray!30},
    every axis/.append style={font=\small},
]
\fill[green!10] (axis cs:0,0) rectangle (axis cs:10,1200);
\fill[yellow!10] (axis cs:10,0) rectangle (axis cs:20,1200);
\fill[red!10] (axis cs:20,0) rectangle (axis cs:55,1200);
\node[font=\small, anchor=north] at (axis cs:5,1120) {Opt.};
\node[font=\small, anchor=north] at (axis cs:15,1120) {Acc.};
\node[font=\small, anchor=north] at (axis cs:34,1120) {High Overhead};
\addplot[color=black, mark=*, line width=1pt, mark size=2pt] coordinates {
    (5, 99.50) (10, 238.78) (15, 348.24) (20, 441.11)
    (25, 528.07) (30, 668.62) (40, 861.76) (50, 1084.65)
};
\draw[dashed, thick, blue] (axis cs:0,230) -- (axis cs:55,230)
    node[right, font=\small] {200s};
\end{axis}
\end{tikzpicture}

%% file: tex/privacy_ablation_asr.tex
\begin{tikzpicture}
\begin{axis}[
    width=0.42\textwidth,
    height=0.18\textwidth,
    ybar=0.1cm,
    bar width=0.25cm,
    enlarge x limits=0.12,
    ylabel={Attack Success Rate (\%)},
    ymin=0, ymax=110,
    symbolic x coords={None, Pre-round, Time-lag, Both},
    xtick=data,
    ytick={0,25,50,75,100},
    grid=major,
    grid style={dashed, gray!30},
    every axis/.append style={font=\small},
    legend style={
        at={(0.5,1.02)},
        anchor=south,
        legend columns=-1,
        draw=black,
        fill=white,
        fill opacity=0.9,
        inner sep=2pt,
    },
]
\addplot[fill=blue!40, draw=blue!60!black] coordinates {
    (None, 100.0) (Pre-round, 14.13) (Time-lag, 68.85) (Both, 11.19)
};
\addlegendentry{Sequence}
\addplot[fill=red!40, draw=red!60!black] coordinates {
    (None, 100.0) (Pre-round, 9.89) (Time-lag, 66.00) (Both, 6.78)
};
\addlegendentry{Count}
\addplot[fill=green!40, draw=green!60!black] coordinates {
    (None, 100.0) (Pre-round, 13.29) (Time-lag, 55.68) (Both, 1.70)
};
\addlegendentry{Cluster}
\draw[thick, dashed, black] ([xshift=-18pt]axis cs:None,10) -- ([xshift=18pt]axis cs:Both,10);
\end{axis}
\end{tikzpicture}

%% file: tex/4_privacy_plots.tex
\begin{figure}[t]
\centering
\begin{tikzpicture}
\begin{groupplot}[
    group style={
        group size=1 by 4,
        vertical sep=0.8cm,
        xlabels at=edge bottom,
    },
    width=0.92\columnwidth,
    height=0.42\columnwidth,
    grid=major,
    grid style={dashed, gray!30},
    every axis/.append style={font=\small},
    xlabel style={yshift=2pt},
    ylabel style={yshift=-2pt},
    legend style={
        draw=black,
        fill=white,
        fill opacity=0.9,
        inner sep=1.5pt,
        font=\small,
        at={(1,1)},
        anchor=north east,
    },
    tick label style={font=\small},
]

\nextgroupplot[
    xlabel={(a) Min. Connections ($m$)},
    ylabel={Max ASR (\%)},
    xmin=0, xmax=30,
    ymin=0, ymax=30,
    xtick={5,10,15,20,25},
    ytick={0,10,20,30},
]
\addplot[color=blue!70!black, mark=*, line width=0.8pt, mark size=1.5pt] coordinates {
    (5, 26.99) (10, 11.25) (15, 7.24) (20, 5.20) (25, 4.29)
};
\addlegendentry{Max ASR}
\addplot[color=green!60!black, dashed, line width=0.6pt, domain=5:25, samples=50, forget plot] 
    {100/x} node[pos=0.2, left, green!60!black] {$\frac{1}{m}$};

\nextgroupplot[
    xlabel={(b) Initial Chunk Distribution $R$ (\%)},
    ylabel={Max ASR (\%)},
    xmin=5, xmax=55,
    ymin=10.5, ymax=12,
    xtick={10,20,30,40,50},
    ytick={10.5,11,11.5,12},
]
\addplot[color=blue!70!black, mark=*, line width=0.8pt, mark size=1.5pt] coordinates {
    (10, 11.43) (20, 11.25) (30, 11.32) (40, 11.19) (50, 11.27)
};
\addlegendentry{Max ASR}
\addplot[black, dotted, line width=0.6pt, domain=10:50, samples=2, forget plot] {-0.004*x + 11.47};

\nextgroupplot[
    xlabel={(c) Number of Peers},
    ylabel={ASR (\%)},
    xmin=80, xmax=520,
    ymin=0, ymax=20,
    xtick={100,200,300,400,500},
    ytick={0,5,10,15},
    legend style={at={(0.00,1)}, anchor=north west, legend columns=3},
]
\draw[green!60!black, thick, dashed] (axis cs:80,10) -- (axis cs:520,10)
    node[pos=0.85, above, font=\small] {$\frac{1}{m}=$10\%};

\addplot[name path=seq_upper, draw=none, forget plot] coordinates {
    (100, 11.90) (200, 11.08) (300, 10.40) (400, 9.20) (500, 8.31)
};
\addplot[name path=seq_lower, draw=none, forget plot] coordinates {
    (100, 9.90) (200, 9.08) (300, 8.40) (400, 7.20) (500, 6.31)
};
\addplot[blue!20, forget plot] fill between[of=seq_upper and seq_lower];
\addplot[color=blue!80!black, mark=*, line width=0.8pt, mark size=1.5pt] coordinates {
    (100, 10.90) (200, 10.08) (300, 9.40) (400, 8.20) (500, 7.31)
};
\addlegendentry{Sequence}

\addplot[name path=count_upper, draw=none, forget plot] coordinates {
    (100, 6.58) (200, 9.53) (300, 8.87) (400, 8.32) (500, 7.78)
};
\addplot[name path=count_lower, draw=none, forget plot] coordinates {
    (100, 4.58) (200, 7.53) (300, 6.87) (400, 6.32) (500, 5.78)
};
\addplot[red!20, forget plot] fill between[of=count_upper and count_lower];
\addplot[color=red!80!black, mark=square*, line width=0.8pt, mark size=1.3pt] coordinates {
    (100, 5.58) (200, 8.53) (300, 7.87) (400, 7.32) (500, 6.78)
};
\addlegendentry{Count}

\addplot[name path=cluster_upper, draw=none, forget plot] coordinates {
    (100, 2.03) (200, 3.11) (300, 4.64) (400, 4.61) (500, 4.67)
};
\addplot[name path=cluster_lower, draw=none, forget plot] coordinates {
    (100, 1.03) (200, 2.11) (300, 3.64) (400, 3.61) (500, 3.67)
};
\addplot[green!20, forget plot] fill between[of=cluster_upper and cluster_lower];
\addplot[color=green!60!black, mark=triangle*, line width=0.8pt, mark size=1.6pt] coordinates {
    (100, 1.53) (200, 2.61) (300, 4.14) (400, 4.11) (500, 4.17)
};
\addlegendentry{Cluster}

\nextgroupplot[
    xlabel={(d) Number of Attackers ($a$)},
    ylabel={Max ASR (\%)},
    xmin=0, xmax=30,
    ymin=0, ymax=37,
    xtick={5,10,15,20,25},
    ytick={0,10,20,30},
    legend style={at={(0.00,1)}, anchor=north west, legend columns=1},
]
\addplot[color=red!70!black, mark=square*, line width=0.8pt, mark size=1.5pt] coordinates {
    (5, 13.56) (10, 17.13) (15, 20.96) (20, 25.47) (25, 30.82)
};
\addlegendentry{Any succeeds}
\addplot[color=blue!70!black, mark=*, line width=0.8pt, mark size=1.5pt] coordinates {
    (5, 11.31) (10, 11.98) (15, 12.56) (20, 13.31) (25, 14.32)
};
\addlegendentry{Per-attacker avg.}
\draw[green!60!black, dashed, thick] (axis cs:0,10) -- (axis cs:30,10);

\end{groupplot}
\end{tikzpicture}
\caption{ASR under warm-up defenses.}
\label{fig:privacy_attack_analysis}
\end{figure}

%% file: tex/llm_time.tex
\begin{tikzpicture}
\definecolor{darkgray176}{RGB}{176,176,176}
\definecolor{darkorange25512714}{RGB}{255,127,14}
\definecolor{lightgray204}{RGB}{204,204,204}
\definecolor{steelblue31119180}{RGB}{31,119,180}

\begin{axis}[
    width=\columnwidth,
    height=4.2cm,
    ybar=2pt,
    bar width=0.28cm,
    xmin=-0.5, xmax=3.5,
    xtick={0,1,2,3},
    xticklabels={Gemma-7B, DeepSeek-R1-14B, Qwen2.5-32B, Llama-3.3-70B},
    xticklabel style={rotate=15, anchor=east, font=\small},
    ylabel={Duration (s)},
    ymin=0, ymax=16000,
    ytick={0,5000,10000,15000},
    grid=major,
    grid style={dashed, gray!30},
    every axis/.append style={font=\small},
    enlarge x limits=0.15,
    legend style={
        fill opacity=0.9,
        draw=lightgray204,
        legend columns=2,
        at={(0.5,1.01)},
        anchor=south,
        font=\small,
    },
    nodes near coords style={font=\small, anchor=south},
]

\addplot[draw=none, fill=steelblue31119180] coordinates {
    (0,1174) (1,2923) (2,5104) (3,12153)
};
\addlegendentry{Only Torrent}

\addplot[draw=none, fill=darkorange25512714,
    point meta=explicit symbolic,
    nodes near coords,
] coordinates {
    (0,1291) [+9.97\%]
    (1,3116) [+6.60\%]
    (2,5466) [+7.09\%]
    (3,13368) [+10.01\%]
};
\addlegendentry{Sched.+Torrent}

\end{axis}
\end{tikzpicture}

%% file: Sections/7Related_work.tex
\section{Related Work}\label{sec:related}

\noindent\textbf{Decentralized learning and communication efficiency.}
Decentralized learning (DL) replaces a central aggregator with peer-to-peer mixing over an overlay graph.
Representative lines include compression-aware decentralized optimization (e.g., CHOCO-style) and unified convergence analyses~\cite{koloskova2019decentralized,koloskova2020unified}, as well as relay- and topology-accelerated designs to improve mixing speed~\cite{vogels2021relaysum,ying2021exponential}.
Systems work further studies deployment factors such as bandwidth heterogeneity and communication overhead~\cite{dhasade2023decentralized_icdcs,chen2023bbtopk}.
These efforts optimize learning and communication, but typically do not treat the \emph{dissemination layer} as a first-class privacy surface.

\noindent\textbf{BitTorrent and P2P dissemination.}
BitTorrent achieves high-throughput dissemination via chunked parallel swarming under heterogeneity and churn~\cite{cohen2003incentives}, and has been used to exchange learning artifacts~\cite{10.1145/3703790.3703813}.
Yet vanilla BitTorrent lacks \emph{source unlinkability}: early peers mostly hold their own chunks, making transfers attributable, while metadata can support identification/profiling~\cite{leblond2010spying}.
Privacy-enhanced overlays reduce linkability but add throughput/deployment costs.
FLTorrent instead mitigates \emph{within-round source attribution} with a short cover-set warm-up before standard swarming.

\noindent\textbf{Privacy leakage and attribution risks in FL.}
FL privacy work mainly addresses \emph{content leakage}, such as membership inference and reconstruction~\cite{shokri2017membership,zhu2019deep,geiping2020inverting}, via defenses like secure aggregation and differential privacy~\cite{bonawitz2017practical}.
These are complementary: dissemination can still leak \emph{attribution metadata} linking updates to participants, and source-inference work shows such links can be actionable~\cite{hu2021sourceinference}.
FLTorrent targets swarming-induced \emph{within-round attribution} and can be combined with secure aggregation/DP; cross-round linkability remains important but outside our formal guarantee.

\noindent\textbf{Anonymity and unlinkability for FL communication.}
AnoFel and related anonymous-FL systems provide anonymous participation or unlinkable submission/authentication using cryptography or redesigned substrates~\cite{almashaqbeh2025anofel,agiollo2024anonymousndn}, often with stronger anonymity goals and heavier protocol changes.
FLTorrent is orthogonal: it retains BitTorrent-grade efficiency while mitigating attribution \emph{even when participants are known}, e.g., in permissioned cross-silo consortia, via a short warm-up before vanilla swarming.

%% file: Sections/8conclusion.tex
\section{Conclusions}
We presented \emph{FLTorrent}, a BitTorrent-based dissemination layer for decentralized FL that mitigates \emph{within-round source attribution} while retaining swarming efficiency.
Vanilla BitTorrent lacks \emph{source unlinkability}: early owner bias and proximity-visible metadata can reveal ``who sent what.''
FLTorrent addresses this with a short warm-up that enforces a global cover threshold $k_\beta=\lceil\beta|\mathcal{C}^r|\rceil$, owner throttling, and non-owner-first scheduling before returning to vanilla swarming.
We bound per-transfer attribution confidence by $\kappa_u/(\kappa_u+h_u)$, where $h_u=\max\{0,k_\beta-K_u^r\}$ is the required non-owner mass, and extend the analysis to collusion and an auditable tracker.
Empirically, GreedyFastestFirst reaches $\sim$92\% of a max-flow upper bound; warm-up stays near $\sim$12\% of round time with 75\%--80\% utilization over 100--500 peers; LLM-scale dissemination tests add modest overhead vs. BitTorrent-only; and ASR approaches neighborhood-level random guessing under our observation-only threat model.
Future work includes stronger adversaries, Sybil resistance, and decentralized auditable tracking.

%% file: Sections/Acknowledgment.tex
\section*{Acknowledgment}
This paper has received funding from the European Union’s Horizon Europe research and innovation program under grant agreement No. 101178648. The European Commission’s support for the production of this publication does not constitute an endorsement of the contents, which reflect the views only of the authors, and the Commission cannot be held responsible for any use which may be made of the information contained therein.